\documentstyle[11pt,epsfig]{article}
\def\baselinestretch{1.3}
\newcommand{\be}{\begin{equation}}
\newcommand{\ee}{\end{equation}}
\newcommand{\bea}{\begin{eqnarray}}
\newcommand{\eea}{\end{eqnarray}}
\newcommand{\pr}{\partial}
\newcommand{\nno}{\nonumber}
\newcommand{\bse}{\begin{subequations}}
\newcommand{\ese}{\end{subequations}}

\newcommand{\bs}{\begin{slide}}
\newcommand{\es}{\end{slide}}
\def\be {\begin{equation}}
\def\ee {\end{equation}}
\def\ba {\begin{eqnarray}}
\def\ea {\end{eqnarray}}
\def\nn {\nonumber}
\def\bc {\begin{center}}
\def\ec {\end{center}}

\def\p  {\pi}

\def\le {\left}
\def\ri {\right}

\def\f {\frac}

\def\no {\noindent}
\def\bi {\begin{itemize}}
\def\ei {\end{itemize}}

\def\ul {\underline}

\def\bc {\begin{center}}
\def\ec {\end{center}}
\begin{document}

\begin{center}
{\Large\bf Aspects of warped braneworld models}\\ [20mm]
Soumitra SenGupta \footnote{E-mail: tpssg@iacs.res.in} \\
{\em Department of Theoretical Physics,\\
Indian Association for the Cultivation of Science\\
Jadavpur, Kolkata- 700 032, India}
\\[20mm]
\end{center}
\begin{abstract}
We review various key issues in connection with the warped braneworld models which provide us with new insights and explanations of 
physical phenomena through interesting geometrical features of such extra dimensional theories. 
Starting from the original Randall-Sundrum two brane models, we have discussed the stability, hierarchy and other 
important issues in connection with such braneworld. The role of higher derivative terms in the bulk for modulus stabilization
has been explained. Implications of the existence of various bulk fields have been discussed and it has been shown
how a warped braneworld model can explain the invisibility of all antisymmetric bulk tensor fields on our brane. 
We have also generalised the model for more than one warped dimensions in the form of a multiply warped spacetime. 
It is shown that such model can offer an explanation to the mass hierarchy among the standard model fermions and the
localization of fermions on the standard model brane with a definite chirality. 
\end{abstract}
\vskip 1cm
\newpage 
\setcounter{footnote}{0}
\def\baselinestretch{1.8}
\section{Background}
Standard model of elementary particles has been extremely 
successful in explaining physical phenomena up to scale close to Tev. 
The background gauge theory 
which refers to the vast disparity between the electroweak scale and the Planck scale 
however gives rise to the the well known hierarchy/fine tuning problem in connection with 
the mass of the only scalar particle in the theory namely Higgs \cite{dreesmartin}.
Higgs mass receives large quadratic quantum correction as,
\ba
\delta m_H^2 &=& \alpha \Lambda^2 \nn
\ea
where $\Lambda$ is the cutoff of the theory i. Planck scale or GUT scale.
To keep $m_H$ within it's allowed value $\sim$ Tev, $\alpha$ must be unnaturally fine tuned to 
$\sim 10^{-32}$, leading to the 'naturalness problem'. Supersymmetry \cite{dreesmartin} may remove  
this large quadratic divergence at the expense of bringing in 
the a large number of superpartners (called sparticles  ) in 
the theory, all of which are so far undetected. This indicates a broken supersymmetry at the present energy scale  
which in turn generates large cosmological constant which is not consistent with it's presently
observed small value.To circumvent this, the local version of supersymmetry called supergravity \cite{nilles} was introduced to
have broken SUSY with zero cosmological constant. 
But the resulting theory is not renormalizable unless one embeds the supergravity model in a more 
fundamental theory like string theory \cite{gsw}.\\
Moreover, what will happen if the signature of supersymmetry is not found near Tev scale in the forthcoming experiments? 
A high scale SUSY (as allowed  in a  Stringy scenario)
will certainly not resolve the hierarchy or the fine tuning problem.
So unless one subscribes to the exotic ideas like landscape or anthropic principle \cite{anthroland} in favour of
fine tuning , we will have to look for some alternative paths to resolve this longstanding issue.\\
Theories with extra spatial dimension(s) is one such  possible alternatives in this direction.
Such theories have 
attracted a lot of attention because of the new geometric approach to solve 
the hierarchy problem. The two most prominent models in this context are 1) ADD model ,proposed by  Arkani-hamed, 
Dimopoulos and Dvali \cite{arkani}, and 
2) RS model, proposed by Randall and Sundrum \cite{lisa}.\\
In ADD model the extra spatial dimension(s) are large and compactified on circles of radii $R$.
The large radius of the extra dimension pulls down the effective higher dimensional
Planck scale i.e the quantum gravity scale.
For two or more extra dimensions, the large radius $R$ can be chosen consistently 
so that the quantum gravity scale $M_d$ at d-dimension and hence the cut-off of the theory $\Lambda$ has the 
desired value $\sim$ Tev.
However in this process it introduces a new hierarchy of length and therefore mass scale in the theory in the form 
of the large radius $R$ ( much larger than the Planck length).\\
In an alternative scenario,
considering one extra spatial dimension Randall and Sundrum proposed a
$5$ dimensional warped geometric model in an anti-deSitter (ADS) bulk spacetime which we describe now.

\section{Randall-Sundrum Model}

In Randall-Sundrum scenario the extra coordinate $y = r \phi$ is compactified on a $S_1/Z_2$ orbifold with
two 3-branes placed at the two orbifold fixed points $\phi = 0,\pi$,
where $r$ is the radius of $S_1$. 
Using $M_{Pl(5)}\equiv M$ the five dimensional action can be written as,
\begin{eqnarray}
S &=& S_{Gravity} + S_{vis} + S_{hid} \nn \\ 
{}\nn \\
\mbox{where,}~~~~S_{Gravity} &=& \int d^4x~r~d{\phi} 
\sqrt{-G}~[ 2M^3R - \underbrace{\Lambda}_{5-dim}]\nn\\
{}\nn\\
S_{vis} &=&  \int d^4x \sqrt{-g_{vis}}~[L_{vis} - V_{vis}] \nn\\
{}\nn\\
S_{hid} &=&  \int d^4x \sqrt{-g_{hid}}~[L_{hid} - V_{hid}] 
\end{eqnarray}
\ul{Metric ansatz:}
\ba
ds^2=e^{-A}~\eta_{\mu\nu}dx^{\mu}dx^{\nu} + 
r^2 d\phi^2
\ea
Warp factor $A(y)$and the brane tensions are found by solving the 5 dimensional
Einstein's equation with orbifolded boundary conditions
\ba
A&=& 2kr\phi \nn\\
{}\nn\\
V_{hid}&=&-V_{vis}= 24 M^3 k 
~~~~~~~~~~~~~~~~~~~~~~
\le[k^2 =\frac{-\Lambda}{24M^3} \ri]    
\ea
The bulk space-time is taken to be anti-desitter  with a negative cosmological constant $\Lambda$.  
\ba
\le( \f{m_H}{m_0}\ri)^2 &=& 
e^{-2A}|_{\phi=\pi} 
= e^{-2kr\p}  \approx (10^{-16})^2 \nn \\
\Rightarrow kr &=& \f{16}{\p}\ln(10)= 11.6279\dots~~\leftarrow \mbox{RS value}
\nn
\ea
with\\
$k \sim M_P$ and $r \sim l_P$\\
So hierarchy problem is resolved without introducing any new scale.

A large hierarchy therefore emerges naturally from a small conformal factor. \\

In this scenario the five dimensional Planck mass $M_5$ is almost equal to the four dimensional Planck mass $M_4$ for
the value of $kr \sim 11.5$ which is required to resolve the hierarchy problem.

\section{Modulus stabilization}

The tiny value ($\sim$ near Planck length) of the modulus $r$ which measures 
the separation of the two branes is associated with the 
vacuum expectation value (VEV) of a massless four-dimensional scalar field 
(modulus field) which has zero potential, so that $r$ is not determined by the 
dynamics of the model. Goldberger and Wise (GW) \cite{gw} proposed to generate such 
a potential classically by introducing a bulk massive scalar field with quartic 
interaction terms localized at the two 3-branes and finally obtained a value 
$kr \sim 12$ by minimizing the potential, without any fine-tuning of parameters. 
In this analysis however the back-reaction of the scalar field on the background geometry was neglected.
Before examining this modulus stabilization issue more critically,
we begin with a brief description of  Goldberger-Wise analysis, which begins with an 
action of the form:

\begin{eqnarray}
S &=& S_{Gravity} + S_{vis} + S_{hid} + S_{scalar}~, \\
\mbox{where,} &&\nonumber \\
S_{Gravity} &=& \int d^4x~r~d{\phi} \sqrt{G}~[ 2M^3R + \Lambda]\\
S_{vis} &=&  \int d^4x \sqrt{-g_{s}}~[L_{s} - V_{s}]\\
S_{hid} &=&  \int d^4x \sqrt{-g_{p}}~[L_{p} - V_{p}]\\
S_{scalar}&=&\frac{1}{2}\int d^4x \int _{-\pi}^{\pi}\sqrt{G}(G^{AB}\partial_{A}
\Phi\partial_{B}\Phi - m^2 \Phi^2) -
\int \sqrt{-g_{p}}\lambda_{p}(\Phi^2-v_{p})^2- \int d^4x \sqrt{-g_{s}}
\lambda_{s}(\Phi^2 - v_{s})^2.
\end{eqnarray}

Here $\Lambda$ is the five dimensional cosmological constant, 
$V_{s}, V_{p}$ are the visible and hidden brane
tensions. \\
$\Phi$ develops a $\phi$-dependent vacuum expectation value, which has to be
determined from the classical equation of motion,
\begin{equation}
\partial_ \phi(e^{-4\sigma}\partial_\phi \Phi)=m^2 r^2 e^{-4\sigma}\Phi
+ 4 ^{-4\sigma}\lambda_{v} r \Phi(\Phi^2 - v_{s}^2)\delta (\phi-\pi)
+ 4 e^{-4\sigma}\lambda_{p} r \Phi(\Phi^2 - v_{p}^2)\delta (\phi)
\label{eqm}
\end{equation}\\
where $\sigma=kr|\phi|$.
In the bulk (where the delta functions are not relevant), the equation has a 
general solution:
\begin{equation} \label{Phi}
\Phi(\phi)=e^{2\sigma}[A e^{\nu\sigma} + B e^{-\nu\sigma}]
\end{equation}
where $\nu=\sqrt{4 + {m^2}/{k^2}}$.                                         
The solution is now plugged back into the original scalar field action 
and integrated over $\phi$ to obtain an effective 4-dimensional potential
for r, of the form:                         
\begin{equation}
V_{\Phi}(r)=k(\nu + 2)A^2 (e^{2\nu kr \pi} -1) + k(\nu - 2)B^2(1-e^{-2\nu kr \pi}) + 
\lambda_{s} e^{-4kr\pi}[\Phi(\pi)^2 -v_{s}^2]^2 + \lambda_{p}[\Phi(0)^2 - v_{p}^2].
\label{pot}
\end{equation}
The coefficients should be determined by matching the delta-function terms at 
the boundaries. The resultant condition
on $A$ and $B$ is obtained as follows:

\begin{eqnarray} 
k[(2+\nu)A + (2-\nu)B] -2\lambda_{p}\Phi(0)[\Phi(0)^2 - v_{p}^2]=0.\label{bd1}{\label{bd1}}\\
k e^{2kr\pi}[(2+\nu)e^{\nu kr\pi}A + (2-\nu)e^{-\nu kr \pi}B] + 
2\lambda_{s}\Phi(\pi)[\Phi(\pi)^2 - v_{s}^2] = 0.
\label{bd2}
\end{eqnarray}

According to \cite{gw} if one considers infinite $\lambda$
limit, one can choose $\Phi(0)=v_{p}$ and $\Phi(\pi)=v_{s}$ as the minimum 
energy configuration. 

Before re-examining the analysis of \cite{gw} for arbitrary value of $\lambda$ ,we first calculate the first and second 
derivative of the potential to determine the exact extremization condition for the modulus without resorting to any approximation
\cite{addmssg}. 
A long but straightforward calculation yields, the first derivative for the potential (assuming
explicit $r$ dependence of $A$ and $B$) as,
\begin{equation} {\label{V'}}
V'_{\Phi}(r)= -4k \pi \lambda_{s} e^{-4kr\pi}(\Phi(\pi)^2 - v_{s}^2)^2
-4k^2\pi[(\nu+2)e^{2\nu kr\pi}A^2 + (2-\nu)e^{-2\nu kr\pi}B^2 + (4-\nu^2)AB].
\end{equation}
where prime denotes differentiation with respect to $r$.

Using the extremization condition for the 
potential $(V'_{\Phi}(r) = 0)$ an
exact form for the second derivative of the potential may be obtained as
\begin{equation}{\label{V''}}
V''_{\Phi}(r)= 4 k^2 \pi \nu [{(2+\nu)AB' + (\nu - 2)BA'}]
\end{equation}
The sign of R.H.S. of the above equation determines  whether the stationary value of the
modulus $r$ is a stable value or not.

Assume that for arbitrary value of $\lambda$ ( not infinity) the boundary value of the scalar field at the two orbifold fixed
points are $\Phi(\phi = 0) = Q_p(r)$ and $\Phi(\phi = \pi) = Q_s(r)$. Now
The undetermined constants $A$ and $B$ in terms of these quantities,
$Q_p$ and  $Q_s$ are given as,
\begin{eqnarray} \label{AB}
A = \frac {Q_s(r) e^{-2 \sigma} - Q_p(r) e^{- \nu \sigma}}{ 2 \sinh(\nu \sigma)} \\
B = \frac {Q_p(r) e^{\nu \sigma} - Q_s(r) e^{- 2 \sigma}}{ 2 \sinh(\nu \sigma)}
\end{eqnarray}
Substituting the expressions for $A$ and $B$ in Eq.\ref{bc} and equ.\ref{V'} under 
stability condition we arrive at ( for $\lambda_s \neq 0$),
\begin{equation} \label{root}
x^2\left( 1 + \frac k {\lambda_s Q_s^2}\right) = \tilde{C}^2,
\end{equation}
where 
\begin{equation}
x = \frac {Q_p}{Q_s} - \frac{2 + \nu}{2 \nu} e ^{(\nu -2) \sigma}
- \frac {\nu -2}{2 \nu} e^{-(\nu+2) \sigma},~~~~~~~\tilde{C} = \left\{\frac{2 + \nu}{2 \nu} e ^{(\nu -2) \sigma} +\frac {\nu -2}{2 \nu} e^{-(\nu+2) \sigma}\right\} C
\end{equation}

Now, it is easy to manipulate the expression given below from the Eq.\ref{root} that is
\begin{equation}{\label{kr}}
 k r = \frac 1 {\pi (\nu -2)} \ln \left[\left(\frac 1 { \frac {2 + \nu}{2 \nu} + 
 \frac {\nu - 2}{2 \nu}
 e^{-2 \nu \sigma}}\right) \left(\frac {Q_p(r)}{Q_s(r)}\right) 
 \left(\frac 1 {1 \pm C \sqrt \frac {\lambda_s 
 Q_s(r)^2}{k + \lambda_s Q_s(r)^2}}\right)
 \right]  
\end{equation}
for the stationary condition i.e $V'_{\Phi}(r) = 0 $. 
Here
\begin{equation}
 C = \sqrt{ 1 - \frac{ \frac 4 {\nu} \{(2+\nu)e^{2 (\nu -2) \sigma} - 
 e^{- 4 \sigma} (4-\nu^2) + (2 - \nu)e^{-2 (\nu + 2) \sigma}\}}{\{(\nu + 2)e^{(\nu -2)\sigma}
 + (\nu - 2)e^{-(\nu +2)\sigma}\}^2}} 
\end{equation}
It may be noted that in the large $k r$ limit, $C \sim \sqrt{\frac {\nu -2 }{\nu +2 }}$. 
The expression for $k r $ in equ.\ref{kr}), is an exact expression for the stationary value of the
modulus $r$ and is valid for any value of the brane coupling constants $\lambda$.

Using the above results, the expression for $V''$ becomes
\begin{equation} 
V_{\Phi}''(r) = - \frac {4 k \pi \nu e^{- 2 \sigma}}  {\sinh{(\nu \sigma)}} 
\left[ \lambda_p(Q_p^2 - v_p^2)Q_p Q_p'
+ \lambda_s(Q_s^2 - v_s^2)Q_s Q_s' + \pi \lambda_p\lambda_s(Q_s^2 - v_s^2)(Q_p^2 - v_p^2)Q_p Q_s
\right]
\label{V''2}
\end{equation}
where 'prime'$\{'\}$ denotes derivative with respect to $r$.

At this point, we want to reiterate that no approximation has been made so far and
all the results are exact.
\vskip .2cm
{ \it \underline{Stability Analysis}}\\

We now re-examine the stability of the modulus $r$.

{\bf Case I:}~~~~~~ $\lambda \rightarrow \infty $

From the expression for the potential Eq.\ref{pot}, the minimum energy configuration
leads to the 
\begin{equation}
Q_s \rightarrow v_s ~~~;~~~ Q_p \rightarrow v_p
\end{equation}
which are constants. This completely agrees with the result of \cite{gw} along with the following new features:

In this limit the expressions for the stationary points become
\begin{equation}
k r = \frac 1 {\pi (\nu -2)} \ln\left[\frac {2 \nu}{2+\nu} \frac {v_p(r)}{v_s(r)} 
\frac 1 {1 \pm \sqrt{\frac {\nu -2 }{\nu +2 }}}
\right] 
\end{equation}

In the infinity limit of the coupling constant
\begin{equation}
Q_s' = 0, ~~~~~~~;~~~~~~~~~ Q_p' = 0
\end{equation}

and
\begin{equation}
V_{\Phi}''(r) = - \frac {4 k \pi \nu e^{- 2 \sigma}}  {\sinh{(\nu \sigma)}} 
\left[ \pi \lambda_p\lambda_s(Q_s^2 - v_s^2)(Q_p^2 - v_p^2)Q_p Q_s
\right]
\end{equation}

Clearly for a wide range of parameter values, the expression
 $\lambda_p (Q_p^2 - v_p^2)$ is negative. In general it can be both 
positive or negative according to the value of $k r$. However, for  
\begin{equation}
k r_{+} = \frac 1 {\pi (\nu -2)} \ln\left[\frac {2 \nu}{2+\nu} \frac {v_p(r)}{v_s(r)} 
\frac 1 {1 + \sqrt{\frac {\nu -2 }{\nu +2 }}}
\right],
\end{equation}
$V''(r) > 0$, that means $k r_{+}$ is a stable point for the minimum of the potential.
So, for the other value of $k r_{-}$, the potential will be maximum.
Clearly no extreme fine tuning of the parameters is required to get the right magnitude 
for $k r$ . 
It is worthwhile to note that the expression for the value of $ k r$ is different
from that of \cite{gw}. This is why if we calculate
the value of $k r $ corresponding to the values of parameters $m/k = 0.2$ and 
$v_p/v_s = 1.5$ given in \cite{gw} we get $k r = 10.846$ which is less than
what have been predicted in \cite{gw} 
For instance, $v_p/v_s = 2.3$ and $\nu = 2.02$ yields $k r = 12.2504$ for 
the minimum of the potential.
The value of $kr = 14.4993$ corresponds to maximum of the potential. 
Clearly these two values are very close to each other.\\ 

{\bf Case II:}~~~~~~ $\lambda$  is finite but very large, 

From the Eq.\ref{bc}, it is clear that if the brane coupling constant is large but finite, the value of the
scalar field $Q_p$ on the Planck brane is lower than $v_p$ and approaches $v_p$ as the value of the brane coupling 
constant tends to infinity.
We calculate the corrections to the boundary scalar field values
in the leading $1/{\lambda}$ order correction. These become
\begin{eqnarray} \label{modQ}
Q_p(r) = v_p + \frac {k v_s}{\lambda_p v_p^2}  \frac {\nu e^{- 2  \sigma}}{4 \sinh{(\nu\sigma)}}
\left[\frac {v_s}{v_p} - \left\{\frac {2 + \nu}{2\nu} e^{(2 -\nu) \sigma} + \frac {\nu - 2}
{2 \nu} e^{(\nu + 2) \sigma}\right\}\right] \nonumber\\
Q_s(r) = v_s + \frac {k v_p} {\lambda_s v_s^2}  \frac {\nu e^{2 \sigma}}{4 \sinh{(\nu\sigma)}}
\left[\frac {v_p}{v_s} - \left\{\frac {2 + \nu}{2\nu} e^{(\nu-2) \sigma} + \frac {\nu - 2}
{2 \nu} e^{-(\nu + 2) \sigma}\right\}\right]
\end{eqnarray}

In this case the modified expression for $kr$ becomes (in the large $k r$ limit),
\begin{equation} \label{krMod}
kr = \frac 1 {\pi (\nu -2)} \ln\left[\frac {2 \nu}{2+\nu} 
\frac n {1 \pm \sqrt{\frac {\nu -2 }{\nu +2 }}(1 - \frac 1 2 q)} \left\{1 + \frac {\nu -2 }
{2\nu} e^{(2 - \nu) k r \pi} \frac {\nu} 8 \frac t n - \frac {\nu} 8 q n \left(n - \frac {\nu + 2 }
{2\nu} e^{(\nu- 2 ) k r \pi}\right)\right\}\right] 
\end{equation}
where, $\nu, n = v_p/v_s, t = k/(\lambda_p v_p^2)$ and $ q = k/(\lambda_s v_s^2)$ are four
parameters of the model.Here also as in the previous case we obtain one minimum as well as one maximum for the potential. 
The positive sign corresponds to the  value of $kr$ which minimizes
the radion potential.

Our analysis therefore shows that the modulus $r$ may be stabilized even for finite but large value of $\lambda$.
The stable value of the modulus in this case once again solves the hierarchy problem without any unnatural fine tuning of the parameters
although the stable value for the modulus differs marginally from that estimated by GW.
We have determined the modified value of $kr$ because of the the finiteness of $\lambda$ ,to the leading order correction 
around the infinite value i.e. in terms of inverse of $\lambda$.  
As stated earlier our exact calculation indicates that there exists simultaneously
a very closely spaced ($\sim$ Planck length$(l_p)$) maximum along with the minimum.This may have interesting consequences 
in a quantum mechanical version of such a model leading to the possibility of tunneling from one radion vev to the other.

The minimum of this effective potential gives 
us the stabilized value of the compactification radius $r_c$. Several other 
works have been done in this direction \cite{luty-sundrum,maru,kogan,ferreira,csaki1,csaki2,ssg}.
However, in the calculation of GW \cite{gw}, the back-reaction of the 
scalar field on the background metric was ignored. Such back-reaction
was included later in ref. \cite{gub} and exact solutions for the background metric 
and the scalar field have been given for some specific class of potentials, motivated by 
five-dimensional gauged supergravity analysis \cite{freedman}.
Assigning appropriate values of the scalar field on the two branes, a stable value of the modulus $r$ was estimated
from the solution of the scalar field. 
Another interesting work is in the context of a scalar field potential with a supersymmetric form, where 
it has been shown that the resulting model is stable\cite{smyth}.\\

We now examine the modulus stabilization by resorting to the usual modulus potential
calculation and it's subsequent minimization for a very general class of bulk scalar action. 
Keeping the non-canonical as well as higher derivative term in the bulk scalar action
we find the general condition for the modulus stabilization for   
back-reacted RS model \cite{lisa}. We exhibit the role of higher derivative terms 
in stabilizing the modulus as well as resolving the gauge hierarchy problem.

\section{Higher derivative terms and stability  \label{general}}

We start with a general action similar to that in our earlier work \cite{dmssgss}. In the 
last few years there have been many models where the presence of a bulk scalar field is
shown to have an
important role in the context of stability issue of brane-world scenario, bulk-brane
cosmological dynamics, higher dimensional black hole solutions and also in many other 
phenomenological issues in particle physics \cite{gw,gub,bulk1,bulk2,other,particle}.
Here we resort to a somewhat general type of self interacting scalar field along 
with the gravity in the bulk in order to analyze the stability of the RS type two-brane
model. We consider the following 5-dimensional bulk action 
\be \label{action}
S ~=~ \int d^5 x ~\sqrt{-g}\left[ - M^3 R ~+~ F(\phi, X) ~-~ V(\phi)\right] ~-~
\int d^4 x ~dy~ \sqrt{- g_a} \delta (y - y_a) \lambda_a (\phi) .
\ee
where $X = \pr_A {\phi} \pr^A {\phi}$, with `$A$' spanning the whole 5-dimensional bulk
spacetime. The index `$a$' runs over the brane locations and the corresponding brane 
potentials are denoted by $\lambda_a$. The scalar field is assumed to be only the function of extra 
spatial coordinate $y$.

Taking the line element in the form
\be \label{metans}
ds^2 ~=~ e^{- 2 A(y)} \eta_{\mu \nu} dx^\mu dx^\nu ~-~ dy^2 , 
\ee
where $\{y\}$ is the extra compact coordinate with radius $r_c$ such that 
$dy^2 = r_c^2 d\theta^2$, $\theta$ being the angular coordinate.
The field equations turn out to be

\bea
F_X \phi'' - 2 F_{XX} ~{\phi'}^2 \phi'' &=& 4 F_X \phi' A' - 
\frac {\pr F_X}{\pr \phi}{\phi'}^2 - \frac 1 2 \left(\frac {\pr F}
{\pr \phi} - \frac {\pr V}{\pr \phi}\right) + \frac 1 2 \sum_a \frac {\pr
\lambda_a(\phi)}{\pr \phi} \delta(y-y_a) \\
{A'}^2 &=& 4 C F_X ~{\phi'}^2 ~+~ 2 C ~\Big[F(X,\phi) ~-~ V(\phi)\Big] \\
A'' &=& 8 C F_X ~{\phi'}^2  ~+~ 4 C \sum_a \lambda_a(\phi) \delta (y - y_a) 
\eea

where 
\be
C ~=~ \frac 1 {24 M^3} ~;~ F_X ~=~ \frac {\pr F(X,\phi)}{\pr X} ~;~ 
F_{XX} ~=~ \frac {\pr^2 F(X,\phi)}{\pr X^2} \nno.
\ee
and prime $\{'\}$ denotes partial differentiation with respect to $y$. Two of the
above equations are independent and the other one automatically follows 
from the energy conservation in the bulk.

The boundary conditions are
\bea \label{bcondition} 
2\left(F_X ~\phi' \right)|_{y=0} ~&=&~ \frac 1 2 
\frac{\pr \lambda_0(\phi_0)}{\pr \phi} ~~~;~~~ 
- 2\left(F_X ~\phi'\right)|_{y= \pi r_c} ~=~ \frac 1 2 
\frac{\pr \lambda_{\pi}(\phi_{\pi})}{\pr \phi} \\  
2 A'(y)|_{y =0} ~&=&~  4 C \lambda_0(\phi_0) ~~;~~
 -2 A'(y)|_{y = \pi r_c} ~=~  4 C \lambda_{\pi}(\phi_{\pi})
\eea
Now, without knowing the solutions of the above equations explicitly, we may 
analyze the stability of the modulus $r_c$, following the mechanism developed by Goldberger 
and Wise \cite{gw}. The brane separation $r_c$ in general is a dynamical variable 
associated with the metric component $g_{55}$. Integrating out the scalar field action 
over the extra coordinate $y$ in the background of the back-reacted five dimensional metric, 
the 4-dimensional effective potential for the brane separation
$r_c$ is obtained as
\bea
V_{eff} (r_c) &=& - 2 \int_0^{r_c\pi} dy ~ e^{- 4 A(y)} \Big[
- M^3 R ~+~ F(X,\phi) ~-~ V(\phi)\Big] 
 +~ e^{- 4 A(0)} \lambda_0 (\phi_0) ~+~ e^{- 4 A(r_c \pi)} 
\lambda_\pi (\phi_\pi) .
\eea

It may be noted that the effective potential is calculated with the warp factor $A(y)$, which takes care
of the full back reaction of the scalar field on the 5D metric through the equations of motion.
Therefore the effective potential for the modulus $r_c$ is 
calculated by integrating out the full action  
in the five dimensional background back-reacted metric. The modulus $r_c$ in general can be a dynamical variable and the
minimum of its effective potential determines the corresponding stable value. The role of bulk scalar field here is 
to stabilize the modulus associated with $g_{55}$ in 
the bulk five dimensional background spacetime. 

Now, using the above two boundary conditions and expression for the
potential from the above equation of motion 
\be
V(\phi) = - \frac 1 {2 C} {A'}^2 + 2 F_X ~{\phi'}^2 ~+~ ~F(X,\phi),
\ee
and also the expression for the Ricci scalar $R = 20 A'(y)^2 - 8 A''(y)$,
one gets the expression for the effective potential as
\bea
V_{eff} &=& - 16 M^3 \left[ A'(0) - A'(r_c \pi) e^{- 4 A(r_c \pi)} \right].
 \eea
 
Now, taking derivative with respect $y_{\pi} = \pi r_c$,
one finds, by the use of the equations of motion, 
the following algebraic
equation 

\be \label{minimum}
\frac {\pr V_{eff}(r_c)}{\pr (\pi r_c)} = 16 M^3 e^{-4 A(y)}
\left[A''(y) - 4 A'(y)^2\right]_{\pi r_c}
\ee

In order to have an extremum for $V_{eff}$ at some value of $r_c$, the right 
hand side of Eq.(\ref{minimum}) must vanish at that (stable) value of $r_c$.
This immediately implies that the value $A''(y)$ must be positive
and equals to the value of $4 A'(y)^2$ at $y = r_c$. Thus for different 
solutions of the warp factor for different bulk scalar actions, the above condition
determines the corresponding stable value of the modulus $r_c$.
  
 Let us now resort to the general solutions of the full set of field equations.

We start with the case of a bulk scalar action with a simple non-canonical kinetic term, 
without any higher derivatives: 
\be \label{spmodel}
F(X,\phi) ~=~ f(\phi)~ X ,
\ee 
where $f(\phi)$ is any well-behaved explicit function of the scalar field $\phi$. Let us assume 
that $f(\phi) = \pr g(\phi)/\pr \phi$, where $g (\phi)$ is another explicit function of $\phi$. 
Then for a specific form of the potential
\be \label{nonpot}
V (\phi) ~=~ \frac 1 {16} \left(\frac {\pr g}{\pr \phi}\right)\left[ \frac{\pr W}{\pr g} \right]^2 ~-~ 
2 C ~W(g(\phi))^2
\ee
it is straightforward to verify, for some $W(g(\phi))$ and $g(\phi)$, that a solution to 
\be \label{soleq1}
\phi' ~=~ \frac 1 4 \frac{\pr W}{\pr g} ~;~~~
A' ~=~ 2 C ~W(g(\phi)) .
\ee
are valid solutions provided 
\be \label{bc}
\left[g(\phi)~\phi'\right]_a ~=~ \frac 1 2 \frac {\pr \lambda_a}
{\pr \phi_a}(\phi_a)~;~~~
\left[A'\right]_a ~=~ 4 C \lambda_a (\phi_a) .
\ee

Now, let us consider a more general case to include higher derivative term such as,
\be \label{genmodel}
F(X,\phi) ~=~  K(\phi) X ~+~ L(\phi) X^2 .
\ee
One of the motivations to consider this type of term in the Lagrangian originates from 
string theory \cite{mukhanov}. The low-energy effective string action contains higher-order 
derivative terms coming from $\alpha'$ and loop corrections, where $\alpha'$ is related to 
the string length scale $\lambda_s$ via the relation $\alpha' ~=~ \lambda_s/{2\pi}$. 

Thus the 4-dimensional Lagrangian involves a non-canonical kinetic term for the scalar field.
With appropriate redefinition of the scalar field we can recast such a term in the Lagrangian as
\be \label{genF}
F(X, \phi) ~=~ f(\phi)[X ~-~ \beta X^2]
\ee
where $\beta$ is a constant parameter (in this case it is equal to unity) and $\{X, \phi\}$ are new 
variables in terms of old variables. This type of action is common in K-essence cosmological 
inflationary models \cite{kessence}, which is equivalent to a scalar action 
in the 5-dimensional bulk with an appropriate potential function. The scalar field is assumed to depend
only on the extra (fifth) dimension $y$.

Now, assuming $f(\phi) = \pr g(\phi)/\pr \phi$, for a specific form of the potential
\be
V(\phi) ~=~ \frac 1 {16} \left(\frac {\pr g}{\pr \phi}\right)\left[ 
\frac {\pr W}{\pr h} \right]^2  + \frac {3 \beta}{256} \left(\frac 
{\pr g}{\pr \phi}\right) \left[\frac {\pr W}{\pr h}\right]^4- 2 C W^2
\ee
and for some arbitrary $W(h(g(\phi)))$ and $g(\phi)$, it is straightforward to 
verify that a solution to 
\be \label{soleq}
\phi' ~=~ \frac 1 4 \frac{\pr W}{\pr h} ~;~~~
A' ~=~ 2 C W(h(g(\phi))) ,
\ee
with the constraint relation
\be \label{relation}
\frac {dh} {dg} ~=~ {1 ~+~ \frac {\beta} 8 \left(\frac {\pr W}{\pr h}\right)^2} ,
\ee 
is also a solution provided we have
\be \label{bc2}
\left[g(\phi)(1 ~- 2 \beta X)~\phi'  \right]_a 
~=~ \frac 1 2 \frac {\pr \lambda_a}{\pr \phi_a}(\phi_a)~;~~~
\left[A'\right]_a ~=~ 4 C \lambda_a (\phi_a) .
\ee
In the $\beta \rightarrow 0$ limit we at once get back the system of equations 
dealing with only the simple non-canonical kinetic term (\ref{spmodel}) in the 
scalar action, as discussed in the first part of this section. If in addition,
$f(\phi) \rightarrow 1/2$, one deals with the usual canonical kinetic term which 
has been discussed in detail in ref.\cite{gub}.

Considering now
\be \label{ansatz}
W(h) ~=~ \frac k {2 C} ~-~ u ~h^2(g(\phi)) ~~~;~~~~ g(\phi) ~=~ \alpha \phi,
\ee
where $k,~ u$ and $\alpha$ are the initial constant parameters of our model with 
their appropriate dimensions, Eq.(\ref{relation}) gives the solution for $h(g(\phi))$:
\be
h(g(\phi)) ~=~ \frac 1 {u \sqrt{\beta/2}} \tan \left(u \sqrt{\beta/2} ~\alpha \phi \right)
\ee
Clearly, in the limit $\beta \rightarrow 0$ we have $h = g$, which corresponds to what 
only the simple non-canonical kinetic term gives us [Eqs.(\ref{nonpot},~\ref{soleq1})]. 

From Eq.(\ref{soleq}) the solution for $\phi$ becomes
\be \label{solphi}
\sin \left(u \sqrt{\beta/2} ~\alpha \phi \right) ~=~ A_0 e^{- u \alpha y/2}
\ee
where $A_0 = \sin \left(u \sqrt{\beta/2} ~\alpha \phi_0 \right)$ 
with $\phi\vert_{y=0} \equiv \phi_0$. The solution for the warp factor $A(y)$ takes 
the form
\be \label{warp}
A (y) ~=~ ky ~-~ \frac {4C} {\beta u^2 \alpha} \ln 
\left({1 - A_0^2 e^{-u \alpha y}}\right) .
\ee
This is the exact form of the back-reacted warp factor where the first term on the 
right hand side is same as that obtained by RS in absence of any scalar field and
the second term is the result of back-reaction due to the bulk scalar. 

In the limit $\beta \rightarrow 0$ and $\alpha =$ constant, 
\be
A (y) ~=~ k y ~+~ 2 C \alpha \phi_0^2 ~e^{-u \alpha y}
\ee
which is the exactly what has been discussed in ref.\cite{gub}.

Following the Goldberger-Wise mechanism \cite{gw}, we now analyze the stability of our
specific model, with $F (X, \phi)$ given by Eq.(\ref{genF})
in the preceding section. From Eq.(\ref{minimum}), we have the extremal condition
in a generic situation:
\be \label{mincond}
4 {A'}^2 ~- A'' ~=~ 0 .
\ee
Now, putting the expressions for the various derivatives of
the metric solution $A$ in the above Eq.\ref{mincond}
one gets, 
\bea
{\cal Q} {\cal Y}^2 + {\cal P} {\cal Y} - \frac {\beta u^2 \alpha k^2}{C} = 0,
\eea
where,
\bea
{\cal Q} = u^2 \alpha^2 \left[ 1 - \frac {16 C}{\beta u^2 \alpha}\right]~~;~~ {\cal P} = u \alpha ( 8 k + u \alpha)~~;~~
{\cal Y} = \frac {A_0^2 e^{ - u \alpha y}}{1 - A_0^2 e^{ - u \alpha y}}
\eea

From the above equation clearly, we can have two different cases.

{\bf Case i}) $ {\cal Q} > 0$, then
the above equation has only one real solution considering the
fact that ${\cal Y} > 0$. The corresponding root is
\bea
 {\cal Y} = \frac 1 {2 {\cal Q}} \left( \sqrt{{\cal P}^2 + \frac {4 {\cal Q} \beta u^2 \alpha k^2}{C}} - {\cal P}
 \right)
 \eea
which gives us the maximum of the potential. So, the point we get
is unstable.

{\bf Case ii})  $ {\cal Q} < 0$, which in turn
says $ \beta << 1$ provided $u$ $\sim$ Plank scale,
then we have two solutions
\bea
{\tilde{\cal Y}}_{\pm} = \frac 1 {2 |{\cal Q}|} \left(  {\cal P} \pm \sqrt{{\cal P}^2 -
\frac {4 |{\cal Q}| \beta u^2 \alpha k^2}{C}}
\right)
\eea
where $|{\cal Q}|$ is the absolute value of ${\cal Q}$. As we have
checked that the larger value of ${\cal Y}  = {\tilde{\cal Y}}_+$ gives
us the stable point for the modulus. So, naturally, the lower
value ${\cal Y}  = {\tilde{\cal Y}}_-$ gives the unstable point.

For these minimum ${\tilde{\cal Y}}_+$ and maximum
${\tilde{\cal Y}}_-$ of the effective radion potential, one
gets respective distance moduli $y_{\pi}^{\pm}$ as
\bea
y_{\pi}^{\pm}  = \frac 1 {u \alpha} log\left[\frac
{\left({\cal P} + 2 |{\cal Q}| \pm \sqrt{{\cal P}^2 -
\frac {4 |{\cal Q}| \beta u^2 \alpha k^2}{C}} \right) A_0^2}
{{\cal P} \pm \sqrt{{\cal P}^2 - \frac {4 |{\cal Q}| \beta u^2
\alpha k^2}{C}} } \right]
\eea
So, by using the above expression for the stable modulus, we get the expression for stable
value of $A(r_s)$
as
\bea
A(r_s) = m~ \ln\left[\frac {(1 + {\tilde {\cal Y}}_+) A_0^2}
{{\tilde {\cal Y}}_+}\right] -
n~\ln\left[\frac 1 {1 +{\tilde {\cal Y}}_+}\right]
\eea
where,
\bea
m = k/(u \alpha)~~~;~~~~n = {4 C}/(u^2 \beta \alpha),
\eea
For proper choice of the values of the parameter, this warp factor can produce the desired
warping from the Planck scale to Tev scale without any unnatural fine tuning.


\section{Bulk tensor fields}

In the brane world model,  the standard model (SM) fields are assumed to lie on a 3-brane, 
while gravity propagates in a 5-dimensional
anti-de Sitter bulk spacetime \cite{brane1} . A natural explanation of such a description comes
from string theory, with the SM fields arising as excitation modes of an open string
whose ends lie on the brane. The graviton, on the other hand, is a closed string excitation
and hence can propagate into the bulk. The massless
graviton mode has a coupling $\sim 1/M_P$ with all matter, 
while the massive modes have enhanced coupling through the warp factor. It not only 
accounts for the observed impact of gravity in our universe but also raises hopes for
new signals in accelerator experiments \cite{rssig}.  However, there are various antisymmetric 
tensor fields which are excitations of a closed string 
and therefore can be expected to lie in the bulk similarly as gravity. The question
is why are the effects of their massless modes less perceptible than the force of gravitation?

Bulk fields other than gravitons have been studied earlier in RS scenarios, starting
from bulk scalars which have been claimed to be required for 
stabilisation of the modulus \cite{gw}.
Bulk gauge fields and fermions have been considered with various phenomenological
implications \cite{davod}. While some of such scenarios are testable 
in accelerator experiments \cite{bulkac}  or observations in the neutrino sector \cite{gross}, 
in general they do not cause any contradiction with our observations so far. 

However, the situation with tensor fields of various ranks (higher than 1) is slightly 
different. It has been already shown  
a closed string excitation rank-2 antisymmetric tensor field such as the Kalb-Ramond 
excitation \cite{kr} can be in the bulk as 
just as the graviton,  
has similar coupling to matter as gravity. Such a field is equivalent 
to torsion in spacetime \cite{ssgpm}, on which the experimental limits are 
quite severe \cite{torlim}. This apparent contradiction has been resolved  
in \cite{ourprl} where it has been shown 
that the zero mode of the antisymmetric tensor field gets an additional exponential
suppression compared to the graviton on the visible brane. This could well be an
explanation of why we see the effect of curvature but not of torsion in the
evolution of the universe. Can we similarly address the effects of other, higher rank, antisymmetric fields which 
occur in the  NS-NS or RR sector of closed string excitations \cite{ gsw}? We explore it now \cite{ourprd}.

For a rank-n antisymmetric tensor gauge field $X_{a{_1}a_{2}...a{n}}$ ,the corresponding field strength tensor has rank-(n+1). 

\begin{equation}
Y_{a{_1}a_{2}...a{n+1}} = \partial_{[a_{n+1}}X_{a{_1}a_{2}...a{n}]}
\end{equation}

Since a spacetime of dimension D admits of a maximum rank D for an antisymmetric tensor,
one can at most have $(n+1) = D$. Thus any antisymmetric tensor field X can have
a maximum rank $D-1$, beyond which it will all have either zero components or will become
an auxiliary field with the field strength tensor vanishing identically.
In absence of any mass term such an auxiliary  field can be eliminated via the equations of motion due
to gauge invariance. 

For rank-5 field strength tensor $Y_{ABCMN}$
of a rank-4 field, one gets two kinds of terms,namely :

\begin{equation}
Y_{ABCMN} = \partial_{[\mu}X_{\nu\alpha\beta y]}
\end{equation}
and
\begin{equation}
Y_{ABCMN} = \partial_{[y}X_{\mu\nu\alpha\beta]} 
\end{equation}

\noindent
The Latin indices denote bulk co-ordinates, the Greek indices
run over the (3+1) Minkowski co-ordinates and $y$ stands for the compact
dimension coordinate. The first class of terms can be removed using the gauge freedom

\begin{equation}
\delta X_{ABCM} = \partial_{[A} \Lambda_{BCM]}
\end{equation}

\noindent 
which allows the use of 10 gauge-fixing conditions for an
antisymmetric $\Lambda_{BCM}$. As a result one can use 

\begin{equation}
X_{\nu\alpha\beta y} = 0
\end{equation}

The second class of terms do not yield any kinetic energy for
$X_{\mu\nu\alpha\beta}$ on the visible brane, and they can 
thus be removed using the equation of motion.In principle,
such an auxiliary field can have an interaction term of the
form $X_{\mu\nu\alpha\beta}B^{\mu\nu}B^{\alpha\beta}$ with second
rank antisymmetric tensor fields. If such terms at all exist,
they will at most result in quartic self-couplings of the
rank-2 field. Thus the rank-4 antisymmetric tensor fields (and
of course those of all higher ranks) have no role to play in the
four-dimensional world in the RS scenario.

Thus all that can matter are the lower rank antisymmetric tensor fields.
The case of a rank-2 field in the bulk (known as the Kalb-Ramond field) has
been already investigated.

Let us now consider the only antisymmetric tensor field of 
a higher rank, surviving on the 3-brane. This is a rank three
tensor $X_{MNA}$, with the corresponding field strength
$Y_{MNAB}$. The action for such a field in 5-dimensions is

\be
S=\int d^5 x\sqrt{-G} Y_{MNAB}Y^{MNAB}
\ee
where $G$ is the determinant of the 5-dimensional metric.
Using the explicit form of the RS metric and taking into 
account the gauge fixing condition $X_{\mu\nu y}=0$, 
one obtains

\be
S_x=\int{d^4 x\int{d\phi[e^{4\sigma}\eta^{\mu\lambda}\eta^{\nu\rho}\eta^{\alpha\gamma}
\eta^{\beta\delta}Y_{\mu\nu\alpha\beta}Y_{\lambda\rho\gamma\delta}+
4 \frac{e^{2\sigma}}{r_c^2}\eta^{\mu\lambda}\eta^{\nu\rho}
\eta^{\alpha\gamma}\partial_{\phi}X_{\mu\nu\alpha}\partial_{\phi}X_{\lambda\rho\gamma} ]}}
\ee

The Kaluza-Klein decomposition of the field X is,

\be
X_{\mu\nu\alpha}(x,\phi)=\Sigma_{n=0}^{\infty}X^n_{\mu\nu\alpha}(x)\frac{\chi^n(\phi)}{\sqrt{r_c}}
\ee
an effective action of the following form can be obtained in terms of the 
projections  $X^n_{\mu\nu\alpha}$ on the visible brane:

\be
S_X=\int{d^4 x\Sigma_n[\eta^{\mu\lambda}\eta^{\nu\rho}\eta^{\alpha\gamma}\eta^{\beta\delta}
Y^n_{\mu\nu\alpha\beta}Y^n_{\lambda\rho\gamma\delta}+ 4 m_n^2 \eta^{\mu\lambda}\eta^{\nu\rho}
\eta^{\alpha\gamma}X^n_{\mu\nu\alpha}X^n _{\lambda\rho\gamma}]}
\ee
where $m_n^2$ is defined through the relation
\be
-\frac{1}{r_c^2}\frac{d}{d\phi}(e^{2\sigma}\frac{d}{d\phi}{\chi^n})=m_n^2\chi^n e^{4\sigma}
\ee
and $\chi^n$ satisfies the orthonormal condition
\be
\int{e^{4\sigma}\chi^m(\phi)\chi^n(\phi) d\phi}=\delta_{mn}
\ee

In terms of $z_n=\frac{m_n}{k}e^{\sigma}$ and 
$x_n=z_n(\pi)= \frac{m_n}{k}e^{kr_c\pi}$,
the solution for the massive modes can be obtained from the equation as,

\be
\chi^n(z_n)=\sqrt{kr_c}\frac{e^\sigma}{e^{kr_c\pi}}\frac{J_1(z_n)}{J_1(x_n)}
\ee

The values of the first few massive modes of the rank-3 antisymmetric tensor field
are listed in Table 1, where we have also shown the masses of the graviton 
as well as the rank-2 antisymmetric Kaluza-Klein modes. It may be noted that
the rank-3 field has higher mass than the remaining two at every order, and, while
the Kalb-Ramond massive modes  can have some signature at, say,
the Large hadron collider (LHC), that of the rank-3 massive tensor field
is likely to be more elusive.

\begin{center}
$$
\begin{array}{|c|c c c c|}
\hline
n  & 1  & 2  & 3  & 4  \\
\hline
m_n^{grav}~(TeV) & 1.66 & 3.04 & 4.40 & 5.77 \\
\hline
m_n^{KR}~(TeV)  & 2.87 & 5.26 & 7.62 & 9.99 \\
\hline
m_n^{X}~(TeV)  & 4.44 & 7.28 & 10.05 & 12.79 \\
\hline
\end{array}
$$ 
\end{center} 


{\em Table 1: The masses of a few low-lying modes of the graviton,
   Kalb-Ramond (KR) and rank-3 antisymmetric tensor (X) fields, 
   for $kr_c=12$ and $k=10^{19}$Gev.}

\vspace{0.2cm}

Finally we examine the massless mode,
whose strength on the brane needs to be compared to that of the graviton
and the rank-2 field. The solution for this mode is given by
\be
\chi_0=-\frac{C_1}{2kr_c}e^{-2\sigma}+C_2
\ee

Requiring the continuity of $\frac{d\chi^0}{d\phi}$ at 
the orbifold fixed point $\phi = \pi$, one obtains $C_1=0$. The 
normalisation condition then gives
\be
\chi^0=\sqrt{2kr_c}e^{-2kr_c\pi}
\ee

Thus the zero mode of the rank-3 antisymmetric tensor field
is suppressed by an additional exponential factor relative to the 
corresponding rank-2 field which already has an exponential suppression
compared to the zero mode of the graviton. Using the same argument as in
reference \cite{ourprl}, one can extend this result into the coupling of
the field X to matter, and show that the interaction with, say,
spin-1/2 fields is suppressed by a factor $e^{-2kr_c\pi}$.
Thus the higher rank antisymmetric
field excitations have progressively insignificant roles to play on the
visible brane, with the fields vanishing identically beyond rank 3 \cite{ourprd}.

Thus  the graviton has a unique role among the
various closed string excitations in a warped geometry. This is because
the intensity of its massless mode on the 3-brane leads to coupling
$\sim 1/M_P$ with matter fields, which is consistent with the part
played by gravity (or more precisely the curvature of spacetime) 
observed in our universe. On the contrary the bulk antisymmetric
tensor fields upto rank-3 can still have non-vanishing zero modes
in four-dimensional spacetime. However their  strength is progressively
diminished for ranks-2 and 3. This may well serve as an explanation of 
why their role in the observable universe on the Tev 3-brane is imperceptible. The masses of the higher modes also 
tend to increase with rank, making them less and less relevant
to accelerator experiments.

\section{Fine tuning problem in presence of  bulk antisymmetric tensor field}    
It has been discussed in the previous section that
apart from graviton there are other massless closed string excitations also which are free to enter the bulk. 
One of such field is the two form of antisymmetric tensor field namely the Kalb-Ramond (KR) field $B_{MN}$ with the corresponding
third rand antisymmetric tensor field strength $H_{MNL}$ such that $H_{MNL} = \partial_{[M} B_{NL]}$. 
Apart from \cite{ourprl} the implications of the presence of such a bulk field in a RS braneworld has already been analyzed in various 
contexts \cite{ssg1}. Here we re-examine the fine tuning problem in connection with the Higgs mass, 
when such a field exists in the bulk \cite{ssg, sauanissg1}\\ 
We begin with the action,
\begin{eqnarray}
\mbox{where,}~~~~S_{Gravity} &=& \int d^4x~r~d{\phi} 
\sqrt{-G}~[ 2M^3R - \underbrace{\Lambda}_{5-d}] \nn\\
{}\nn\\
S_{vis} &=&  \int d^4x \sqrt{-g_{vis}}~[L_{vis} - V_{vis}] \nn\\
{}\nn\\
S_{hid} &=&  \int d^4x \sqrt{-g_{hid}}~[L_{hid} - V_{hid}] \nn \\
{}\nn\\
S_{KR} &=& \int d^4x~r~d{\phi}\sqrt{-G}~[ H_{MNL} H^{MNL}]  
\end{eqnarray}
We also consider the warped metric ansatz:
$ ds^2=e^{-A}~\eta_{\mu\nu}dx^{\mu}dx^{\nu} + 
r^2 d\phi^2~\leftarrow~\mbox{extra dim}~ $

Solving the five dimensional Einstein's equation, the solution for the warp factor turns out to be,
\ba
e^{-A}&=&\frac{\sqrt{b}}{2kr}\cosh{(2kr\phi+2krc)}~ 
\nn\\
{}\nn\\
{}\nn\\
\frac{2kr}{\sqrt{b}}&=&\cosh(2krc)~~~,~~~~\mbox{such that}~~ A(0)=1  \nn\\
{}\nn\\
{}\nn\\
c &=& - \frac{1}{2kr} \tanh^{-1} \left( \frac{V_{hid} }{24 M^3 k} \right)
=  -\pi + \frac{1}{2kr} \tanh^{-1} \left( \frac{V_{vis} }{24 M^3 k} \right)~
\ea
Here $b$ measures the KR Energy density.

The scalar mass warping is now given by,

\ba
 \le(\f{m_H}{m_0}\ri)^2 &=& 
e^{-2A}|_{\phi=\pi} 
=
\frac{\sqrt{b}}{2kr}\cosh\left[2kr\pi + \cosh^{-1} \frac{2kr}{\sqrt{b}}\right]
\nn\\
{}\nn\\
&=&
\left[
\cosh\left(2 kr\pi\right) - \sinh\left( 2kr\pi\right)
\sqrt{1- \frac{b}{(2kr)^2}}
~\right]
\nn\\
{}\nn\\
&\approx& (10^{-16})^2 \nn \\
{}\nn
\ea
Inverting the above expression we obtain,
\ba
b = (2kr)^2
\left[1 -
\left(  
\coth(2kr \pi) - (m_H/m_0)^2 {\rm cosech}(2kr\pi)
\right)^2 
\right]~
\ea
How large $b$ can be to obtain the desired warping from Planck scale to Tev scale? \\

\no
$\log|b|$ vs $kr$, for $\f{m_H}{m_0}=10^{-16}$\\
We thus have,\\
For $kr$ same as RS value, $b=0$ \\
For $kr<$ RS value, $b<0$ \\
and for $kr>$ RS value, $b>0$

But the maximum possible value for $b$ in this entire range turns out to be,
\ba
b_{max}=10^{-62}
\ea
This indicates extreme fine tuning of $b$ ! 
Thus in order to avoid the fine tuning $\sim 10^{-32}$ of the scalar (Higgs)
mass the warped geometry model was proposed to achieve the desired warping from Planck scale to Tev scale geometrically.
But our result indicates that such geometric warping can be achieved only if the energy density of the bulk
KR field is fine tuned $\sim 10^{-62}$. \\
We have generalized our work by including the dilaton field ( another massless string excitation) also in
the bulk \cite{sauanissg2}. Once again our result revealed that an unnatural fine tuning of the energy density parameter of the
KR field is necessary to achieve the desired warping from the Planck scale to the Tev scale.\\
Thus in the backdrop of a string inspired model the fine tuning problem reappears in a new guise.\\


\section{Generalization to six dimension: Fermion mass splitting}
We now generalize the Randall-Sundrum warped brane model to a spacetime with more than one warped dimensions
\cite{rshigh,soudebchou}.
In a six dimensional warped space-time
we consider a doubly warped space-time as
$M^{1,5} \rightarrow [M^{1,3} \times S^1/Z_2] \times S^1/Z_2$ \cite{soudebchou}.
The non-compact directions would be denoted by $x^\mu \,
(\mu = 0..3)$ and the orbifolded directions by the angular
coordinates $y$ and $z$ with $R_y$ and $r_z$ as
respective moduli.
Four 4-branes are placed at the orbifold fixed points: \\
$y = 0 ,\pi$ and $z = 0,\pi$ with appropriate brane tensions.
Four 3-branes appear at the four intersection region of these 4-branes.
We also consider an ADS bulk with a negative cosmological constant
$\Lambda$. We thus have a brane-box like space-time.

The six dimensional warped metric ansatz:
\ba
ds^2 = b^2(z)[a^2(y)\eta_{\mu\nu}dx^{\mu}dx^{\nu} + R^2_y dy^2] + r^2_z dz^2 \nn
\ea
The total bulk-brane action is thus given by,
\ba
S & = & S_6 + S_5 + S_4 \nn\\
S_6 & = & \int {d^4 x} \, {d y} \, {d z} \, 
          \sqrt{-g_6} \; \left(R_6 - \Lambda \right) \nn\\ 
S_5 & = & \int {d^4 x} \, {d y} \, {d z} \, 
           \left[ V_1 \, \delta(y) + V_2 \, \delta( y - \pi) \right] \nn\\
& + & \int {d^4 x} \, {d y} \, {d z} \, 
           \left[ V_3 \, \delta(z) + V_4 \, \delta(z - \pi) \right] \nn\\
S_4 & = & \int d^4 x \sqrt{-g_{vis}}[{\cal L} - \hat V]  
\ea

In general $V_1, V_2$ are functions of $z$ while $V_3, V_4$ are 
functions of $y$.
The intersecting 4-branes give rise to 3-branes
located at,
$(y, z) = (0,0), (0, \pi), (\pi, 0), (\pi, \pi)$.\\
Substituting the metric in six dimensional Einstein's equation the solutions are:
\ba
a(y) & = & \exp(-c \, y) \nn\\ 
b(z) &=& \frac{\cosh(k \, z)}{\cosh(k \, \pi)} 
\ea
Minimum warping at the 3-brane located at $y = 0, z =\pi$ . Maximum warping at the
3-brane located at $y = \pi, z = 0$. 

Here
\ba
c & \equiv & \frac{R_y \, k}{ r_z \, \cosh(k \, \pi)} \nn\\
k & \equiv &  r_z \, \sqrt{\frac{-\Lambda}{10 \, M^4}}
\ea
Using the orbifolded boundary condition the full
metric thus takes the form,\\
\ba
ds^2 & = & \frac{\cosh^2(k \, z)}{\cosh^2 (k \, \pi)} \,
  \left[ \exp\left(- 2 \, c \, |y| \right)
                \, \eta_{\mu \nu} \, d x^\mu \, d x^\nu  
     + R_y^2 \, d y^2 \right] \nn\\ 
&+& r_z^2 \, d z^2 \ 
\ea
Also the $4+1$ brane tensions become coordinate dependent and are given as,\\
\ba
V_1(z) & = &-V_2(z) = 8M^2 \sqrt{\frac{-\Lambda}{10}} sech(kz) \nn\\ 
V_3(y) & = & 0 \nn \\
V_4(y) & = &-\frac{8 M^4 k}{r_z} \tanh(k \pi)  
\ea
We therefore observe that,
the 3 branes appear at the intersection of the various 4 branes.
There are 4 such 3 branes located at 
$(y, z) = (0,0), (0, \pi), (\pi, 0), (\pi, \pi)$.
The metric on the 3-brane located at $(y = 0, z = \pi)$ suffers no warping.
So it is identified with the Planck brane
Similarly we identify the SM brane with the one at $y= \pi, z = 0$.
Planck scale mass $m_0$ is warped to 
\ba 
m = m_0 \,\frac{r_z \, c}{R_y \, k } \, \exp(-\pi \, c) 
          = m_0 \,  \frac{\exp(-\pi \, c)}{\cosh (k \, \pi)} 
\ea

An interesting picture emerges from this.
To have substantial warping in the $z$-direction (from $z = 0$ to $z = \pi$),
$k \, \pi$ must be substantial, i.e. of same order of magnitude
as in the usual RS case\\
But $c$ is given by 
\ba
c & \equiv & \frac{R_y \, k}{ r_z \, \cosh(k \, \pi)} 
\ea
This immediately means that $c$ must be small for $r_z \sim R_y$\\
Thus we cannot have a large warping in
$y$-direction as well if we want to avoid a new and undesirable
hierarchy between the moduli.
Thus of the two branes located at $y,z =0,0$ and $y,z = \pi ,\pi$,
one must have a natural mass scale close to the Planck scale,
while for the other it is close to the TeV scale.
This resembles to the fine structure splitting of energy levels.

If we repeat this calculation for a seven dimensional
warped space-time , we similarly find, 
eight 3- branes appearing from the intersection of the hyper-surfaces.
Four of these are closed to Tev scale and  
four others are close to Planck scale.
Thus increasing the number of warping in space-time results into
two clusters of branes around Planck and Tev scale.

An interesting phenomenological consequences like mass splitting of the standard model fermions
on the brane follows from this.
The SM-like fields in each of these 3-branes will have apparent mass-scales (on each brane) close to TeV 
with some splitting between them.
To understand this , imagine the SM fermions being defined by 5-dimensional fields 
with $x^\mu$ and $y$ dependence , restricted to the 4-brane say at $z = 0$ which now defines the 
``bulk'' for these fields.
This 4-brane also intersects two other 4-branes at $y = 0$ and $y = \pi$ respectively.
If the major warping has occurred in the $z$-direction, then
the natural mass scale of these fields is still ${\cal O}({\rm TeV})$.
The presence of a $y$-dependence leads to a
non-trivial bulk wave-function.
This,
in turn, changes the overlap of the fermion wave-function with that of
a scalar located on the 3-brane and thus the effective Yukawa
coupling.
The slightly differing interactions on the distant 3-brane
located at two different values of $y$ would result in a hierarchy amongst the effective Yukawa couplings and the fermion masses
on the 3-brane.
Adjusting the parameters suitably the observed mass splitting among the 
different generations of fermions can be explained \cite{fermionmass}. 
It has also been shown that in this model the coordinate dependent brane tension, which is equivalent to a scalar field 
distribution on the 4-brane localizes the left chiral fermionic mode on our 3-brane. Thus the fermion localization
problem is automatically resolved in such a multiply warped spacetime\cite{rkjmssg}. 

\section{Conclusions}
The RS model is known to solve the hierarchy problem with gravity in the bulk. In this review we have discussed
various implications and possible extensions of such warped braneworld model. The issue of modulus stabilization has been explored
in details with higher derivative terms present in the bulk. The role of various bulk antisymmetric tensor fields have been discussed.
Such studies are specially relevant in the context of string inspired models. It is shown that RS model offers a natural explanation
of invisibility of all the antisymmetric tensor fields on our brane and thereby making the status of the  
closed string symmetric tensor excitation namely gravity  quite distinct from the 
antisymmetric closed string excitations. Finally a multiply warped geometric model is proposed as a natural extension to RS model
in six and higher dimensions. Such model is shown to offer a possible  explanation of the standard model fermion mass hierarchy, a
proper resolution of which has been eluding us for a long time. Moreover the consistency 
requirement of such a model leads to the localization of fermions on Tev brane with definite chirality.
Thus a multiply warped spacetime offers a mechanism to obtain chiral fermions in our universe.



\end{document}